%% file: main.tex
\documentclass[11pt, a4paper, logo, copyright, nonumbering]{antgroup}
\usepackage[square, numbers]{natbib}
\usepackage{dblfloatfix}
\usepackage{ulem}
\usepackage{caption}
\usepackage{dramatist}
\usepackage{xspace}
\usepackage{pifont}
\usepackage{multirow}

\usepackage{xltabular}
\usepackage{longtable}
\usepackage{hyperref}
\usepackage{amsfonts}
\usepackage{amsmath}
\usepackage{amssymb}
\usepackage{lineno}
\usepackage{multirow}
\usepackage{adjustbox}

\usepackage[bottom]{footmisc}

\usepackage{CJKutf8}
\usepackage{subfigure}
\usepackage{setspace}

\usepackage{dsfont}
\usepackage{array}
\usepackage{tabularx}
\usepackage{subfigure}
\usepackage{xcolor}
\usepackage{tabularx}
\usepackage{booktabs}

\usepackage{lipsum}
\usepackage{multicol}
\usepackage{authblk}
\usepackage{wrapfig}
\usepackage[most,skins,theorems]{tcolorbox}
\usepackage{array}
\usepackage{pifont}
\definecolor{customgray}{RGB}{230,230,230}

\tcbset{
  aibox/.style={
    width=\linewidth,
    top=8pt,
    bottom=4pt,
    colback=gray!10!white,
    colframe=gray!50!black,
    colbacktitle=gray!70!black,
    enhanced,
    center,
    attach boxed title to top left={yshift=-0.1in,xshift=0.15in},
    boxed title style={boxrule=0pt,colframe=white,},
  }
}
\newtcolorbox{AIbox}[2][]{aibox,title=#2,#1}

\makeatletter
\def\@BTrule[#1]{%
  \ifx\longtable\undefined
    \let\@BTswitch\@BTnormal
  \else\ifx\hline\LT@hline
    \nobreak
    \let\@BTswitch\@BLTrule
  \else
     \let\@BTswitch\@BTnormal
  \fi\fi
  \global\@thisrulewidth=#1\relax
  \ifnum\@thisruleclass=\tw@\vskip\@aboverulesep\else
  \ifnum\@lastruleclass=\z@\vskip\@aboverulesep\else
  \ifnum\@lastruleclass=\@ne\vskip\doublerulesep\fi\fi\fi
  \@BTswitch}
\makeatother

\addto\extrasenglish{
}

 {\begin{list}{}%
         {\setlength{\leftmargin}{#1}}%
         \item[]%
 }
 {\end{list}}

\reportnumber{001} %

\renewcommand{\today}{}

\title{\centering Agentar-DeepFinance-100K: A Large-Scale Financial Dataset via Systematic Chain-of-Thought Synthesis Optimization}

\author{
Xiaoke Zhao$^{*}$,
Zhaowen Zhou$^{*}$,
Lin Chen$^{*}$,
Lihong Wang,
Zhiyi Huang,\ \ \ \ \ \ \ \ \ \ \ \ \ \ \ \ \ \ \ \ \ \ \ \ \ \ \ \ \ \ \ \ \ 
Kaiyuan Zheng,
Yanjun Zheng,
Xiyang Du,
Longfei Liao,
Jiawei Liu, \ \ \ \ \ \ \ \ \ \ \ \ \ \ \ \ \ \ \ \ \ \ \ \ \ \ \ \ \ \ \ \ \ \ \ \ \ \ \ \ \ \ \ \ \  \ \ \ \ \ \ \ \ \ \ \ \ \ \ \ \ \ \ \ \ \ \ \ \ \ \
Xiang Qi,
Bo Zhang$^{\dag}$,
Peng Zhang,
Wei Wang,
and Zhe Li$^{\dag}$
\\
\vspace{-6pt}
Ant Digital Technologies, Ant Group
\vspace{-6pt}
}

\footnotetext[1]{Equal contribution.}
\footnotetext[2]{Corresponding to \{boyuan.zb, lizhe.lz\}@antgroup.com}

\input{commands.tex}

\begin{abstract}
Recent advancements in large language models (LLMs) have demonstrated remarkable general reasoning capabilities, holding significant potential for applications in the financial domain, a field that requires robust and reliable reasoning. It has been demonstrated that distilling high-quality chain-of-thought (CoT) rationales from advanced general reasoning models offers a promising and efficient path to the financial reasoning model. However, existing CoT synthesis methods suffer from shallow CoT sampling, leaving the question of how to construct a well-designed knowledge space for finance reasoning unexplored. In this paper, we present \textbf{Agentar-DeepFinance-100K }, a large-scale financial reasoning dataset characterized by its systematic CoT synthesis optimization. We first introduce a comprehensive CoT synthesis pipeline featuring Multi-perspective Knowledge Extraction (MKE) and Self-Corrective Rewriting (SCR) to generate exhaustive and deep financial reasoning trajectories. Furthermore, a systematic investigation, termed CoT Cube, is conducted to analyze critical factors that influence CoT effectiveness, such as necessity, length and synthesizer, yielding valuable insights for high-quality financial CoT construction. Experiments demonstrate that models trained on our Agentar-DeepFinance-100K  achieve significant improvements on financial benchmarks. We publicly release Agentar-DeepFinance-100K , hoping to advance the research in financial reasoning models. The project page is available at \href{https://github.com/antgroup/Agentar-DeepFinance-100K}{https://github.com/antgroup/Agentar-DeepFinance-100K}.

\end{abstract}

\begin{document}
\maketitle

\section{Introduction}

\begin{figure*}[htbp]
    \vspace{-20pt}
    \centering
    \includegraphics[width=\linewidth]{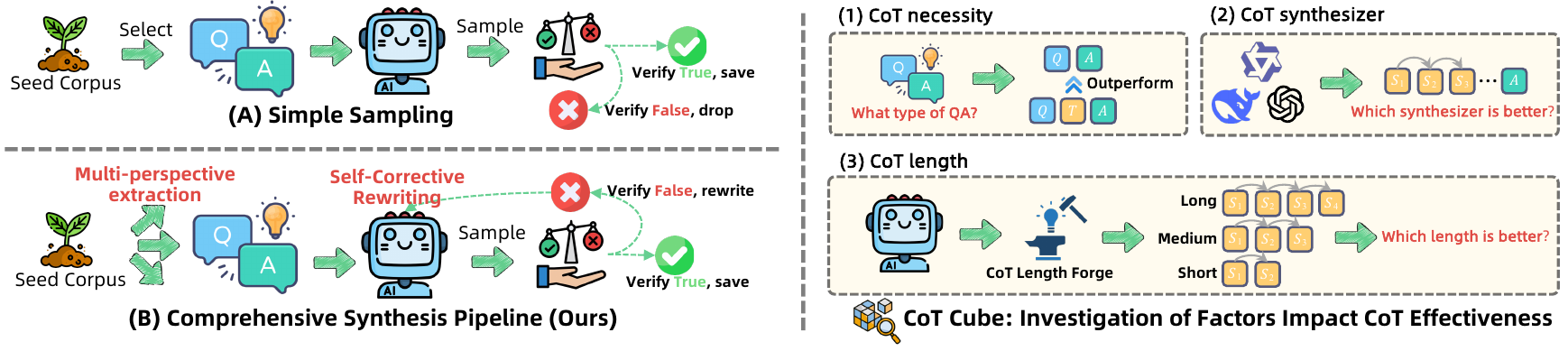}
    \vspace{-20pt}
    \caption{Overview of our proposed CoT synthesis pipeline (left) and systematic investigation of factors that impact CoT effectiveness (right).}
    \vspace{-15pt}
    \label{fig:intro}
\end{figure*}

\noindent Pre-trained on extensive corpora of world knowledge, large language models (LLMs) have showcased impressive generalization capabilities, marking a significant milestone on the way to artificial general intelligence (AGI). When further fine-tuned on domain-specific data, financial LLMs~\cite{xie2023pixiu, wang2023fingpt, yang2023investlm} also exhibit substantial potential in tasks such as financial sentiment analysis~\cite{araci2019finbert}, financial time series~\cite{yu2023temporal} and quant investment~\cite{kou2024automate}. Despite these advancements, such models still face significant challenges in solving financial problems that demand complex reasoning. Encouragingly, recent progress in general reasoning models, such as OpenAI o1~\cite{openaio1}, DeepSeek-R1~\cite{guo2025deepseek} and QwQ~\cite{qwen2.5}, has pushed beyond the limitations of traditional LLMs by explicitly modeling chain-of-thoughts (CoTs). These developments suggest a promising and efficient approach for building financial LLMs, i.e., distilling high-quality financial CoTs from general reasoning models.

Recent works~\cite{liu2025fin,zhu2025dianjin} have demonstrated that synthesizing financial CoTs from general reasoning models can significantly enhance financial reasoning capabilities. However, as illustrated in Figure~\ref{fig:intro} (A), existing methods for CoT synthesis primarily involve directly sampling CoTs from reasoning models for selected question-answer (QA) pairs, followed by rejection sampling for quality filtering. These approaches often result in shallow financial reasoning knowledge extraction, lacking the comprehensive and challenging knowledge required for robust financial reasoning. Additionally, synthesized CoTs are rarely subjected to systematic examinations to identify critical factors (e.g., length) that influence their effectiveness in reasoning enhancement. Without such investigation, it remains unclear how CoTs can be optimized to effectively drive improvements in financial reasoning.

To this end, this paper introduces a comprehensive pipeline for synthesizing CoTs, as shown in Figure~\ref{fig:intro} (B). Beginning with the curated seed corpora, our methodology integrates the Multi-perspective Knowledge Extraction (MKE) strategy and the Self-Corrective Rewriting (SCR) mechanism for CoT sampling. On the one hand, this pipeline demonstrates enhanced capabilities in extracting exhaustive domain knowledge for financial reasoning. On the other hand, it enables the sampling of more complex and challenging questions, leading to more intricate reasoning trajectories. Furthermore, a systematic investigation is conducted regarding critical factors influencing CoT effectiveness, such as necessity, length and synthesizer. As shown in Figure~\ref{fig:intro}, we named it CoT Cube. Empirical evaluations of these components provide valuable insights into CoT-driven optimization of financial reasoning models and offer empirical guidelines for constructing high-performance training datasets.

Employing this method, we present \textbf{Agentar-DeepFinance-100K }, a large-scale financial dataset designed to push the boundary of financial reasoning models. Beyond the synthesizing CoTs, Agentar-DeepFinance-100K  also incorporates multi-dimensional metadata annotation, such as complexity, quality and task type. Agentar-DeepFinance-100K is distinguished from existing financial reasoning datasets by its systematic CoT synthesis optimization. This includes building a comprehensive knowledge space for financial reasoning through our pipeline and refining CoT generation through the exploration of CoT Cube. Furthermore, to bridge the disparity between training data and real-world financial interactions, we additionally curated a dataset annotated by financial experts to reflect the financial capabilities required in real-world scenarios.

The remainder of this report is organized as follows. Section~\ref{sec:overview} provides an overview of our proposed  Agentar-DeepFinance-100K . Section~\ref{sec:pipeline} details our comprehensive CoT synthesis pipeline. Section~\ref{sec:exp} presents an in-depth investigation into the factors influencing the effectiveness of CoT and the ablation study for the CoT synthesis pipeline. Section~\ref{sec:related} discusses related works with this report. Section~\ref{sec:conclusion} summarizes the technical contributions and future works.

\section{Dataset Overview}
\label{sec:overview}

\begin{figure*}[htbp]
    \centering
    \vspace{-10pt}
    \includegraphics[width=\linewidth]{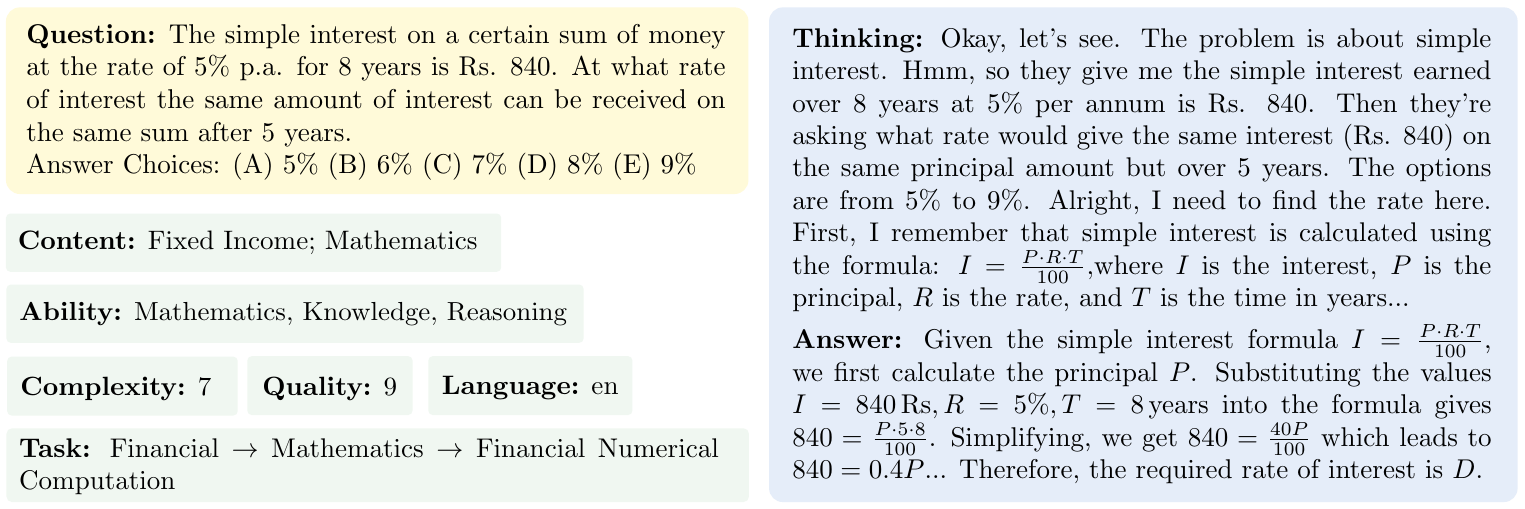}
    \vspace{-10pt}
    \caption{Illustration of the format of Agentar-DeepFinance-100K , which comprises three components: (1) the question, (2) the solution, including CoT and the final answer and (3) metadata, which encompasses multi-dimensional annotations such as complexity, quality and language.}
    \label{fig:dataset_meta}
\end{figure*}

\noindent \textbf{Dataset Format.}
As illustrated in Figure~\ref{fig:dataset_meta}, each sample in the dataset is accompanied by rich metadata in addition to the question and solution. This metadata is annotated using LLMs through a carefully designed manual prompt. Annotating this metadata not only provides a comprehensive understanding of the dataset but also can serve as a basis for subsequent experiments. Specifically, the metadata encompasses the following dimensions:
\begin{itemize}
\item \textbf{Content}: A domain-specific classification of financial topics (e.g., stock).

\item \textbf{Ability}: Classification of the sample’s capability, including language, reasoning, knowledge, mathematics, code, instruction following and agents.

\item \textbf{Complexity}: Numerical score (1-10) represents the difficulty by considering the knowledge depth, multi-step reasoning requirements and instruction compliance challenges.

\item \textbf{Quality}: Score (1-10) evaluating response accuracy, completeness and procedural clarity against financial domain standards. Samples with a quality rating below 8 are excluded from the dataset.

\item \textbf{Language}: Identification of the language used in the sample, such as Chinese (zh) and English (en).

\item \textbf{Task}: A hierarchical decomposition of the task (e.g., Text Creation $\rightarrow$ Marketing Copy Generation).
\end{itemize}

\begin{table}[htbp]
\centering
\begin{tabular}{lrrrrr}
\toprule[1pt]
Dataset                & $N_{\text{Q}}$   & $N_{\text{R}}$ &  $N_{\text{A}}$ & \#Samples & Proportion \\ \midrule
\rowcolor{customgray} \multicolumn{6}{c}{Open-source Dataset}                                                                       \\ \midrule
FinCorpus              & 86.25                & 1775.09           & 330.61               &  61,574      &  $53.64\%$          \\
FinCUGE                & 99.17                & 498.44           & 232.40             &  18,527       &  $16.14\%$          \\
Finance-Instruct-500K  & 151.49               & 920.85           & 558.15             &  10,343      &  $9.01\%$          \\
FinQA                  & 1051.05              & 1928.63         & 189.91              &  5,012        &  $4.37\%$          \\
FinancialData          & 14.85                & 896.85          & 803.57               &  4,047        &  $3.53\%$          \\
Quant-Trading-Instruct & 65.97                & 2665.40         & 903.90            &  89           &  $0.08\%$          \\ \midrule
\rowcolor{customgray} \multicolumn{6}{c}{In-house Dataset}                                                                          \\ \midrule
FinCAS                 & 69.58                & 315.24          & 1.03               &  15,190       &  $13.23\%$          \\ 
FinRA                  &   -                  &  -              &  -                 &  -        &  -          \\
\bottomrule[1pt]
\end{tabular}
\caption{Overview of Agentar-DeepFinance-100K  data sources.}
\label{tab:data_sources}
\end{table}

\noindent \textbf{Data Source.}
The Agentar-DeepFinance-100K  dataset is constructed from two primary sources: 99K samples from open-source data and 16K samples from our in-house data. For the open-source portion, extensive QA pairs were gathered from publicly available datasets, including FinCorpus~\cite{FinCorpus}, Finance-Instruct-500K~\cite{josephgflowers2025financeinstruct}, FinCUGE~\cite{lu2023bbt}, FinQA~\cite{chen2021finqa}, FinancialData~\cite{FinancialData} and Quant-Trading-Instruct~\cite{Quant-Trading-Instruct}. These datasets underwent rigorous filtering to ensure quality and relevance to financial contexts, and instances overlapping with test sets in the benchmarks were systematically removed. The in-house data consists of two primary components: financial real annotations (FinRA) and content related to financial compliance and security (FinCAS). Table~\ref{tab:data_sources} summarizes the data sources, presenting some statistics such as average question length $N_\text{Q}$, reasoning trajectory length $N_\text{R}$ and answer length $N_\text{A}$. Reasoning trajectories were generated using QwQ-Plus.

\begin{figure*}[t]
\centering
\begin{minipage}{0.41\columnwidth}
\centering
\includegraphics[width=\linewidth]{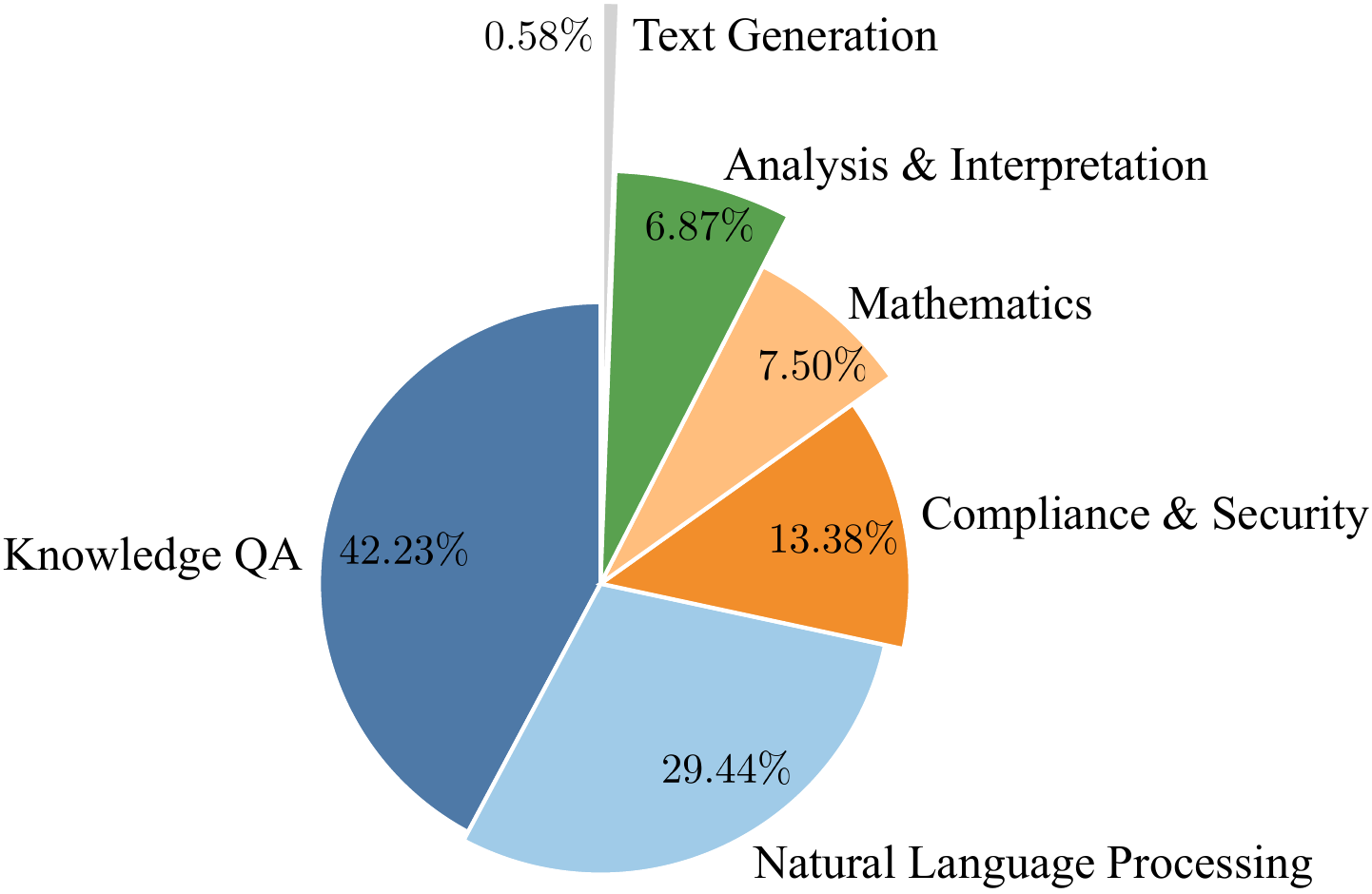}
\vspace{-15pt}
\caption{Task composition.}
\label{fig:task_composition}
\end{minipage}
\hfill
\begin{minipage}{0.56\columnwidth}
\centering
\includegraphics[width=\linewidth]{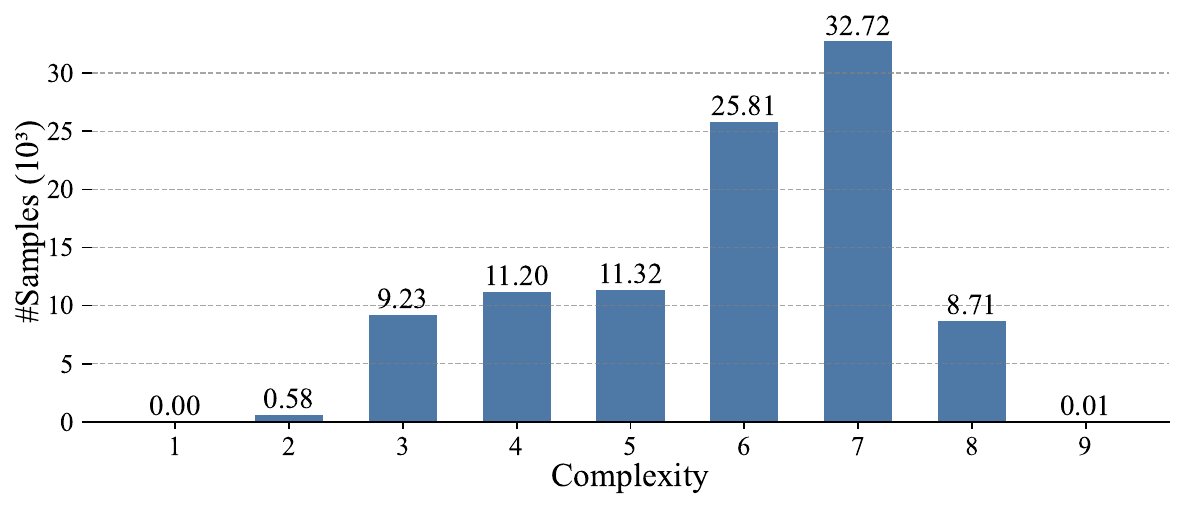}
\vspace{-20pt}
\caption{Complexity distribution.}
\label{fig:complexity}
\end{minipage}
\end{figure*}

\noindent \textbf{Task Diversity Analysis.}
Figure~\ref{fig:task_composition} presents the task composition within our Agentar-DeepFinance-100K , which encompasses six primary task domains: Knowledge QA, Natural Language Processing (NLP), Analysis \& Interpretation, Math, Compliance \& Security and Text Generation. The task distribution exhibits a significant concentration, with Knowledge QA and NLP collectively accounting for over $70\%$ of the dataset. In contrast, Text Generation tasks represent the smallest proportion, accounting for less than $1\%$. This distribution pattern aligns with observed user query tendencies in the financial domain, where interactions predominantly involve straightforward queries related to basic financial knowledge and information.

\noindent \textbf{Distribution of Complexity.}
Figure~\ref{fig:complexity} illustrates the distribution of complexity scores assigned to the Agentar-DeepFinance-100K , with scores ranging from 1 to 10. These complexity scores are generated by a large language model using a carefully crafted prompt designed to capture the relative complexity in solving the problems, rather than directly correlating with the difficulty of the tasks themselves. The results indicate that the complexity scores within the dataset are primarily concentrated in the range of 6 to 7, corresponding to medium-level questions. Additionally, a significant proportion of questions are scored between 3 and 5, which reflects lower levels of complexity. Furthermore, a smaller subset of the dataset includes questions with complexity scores exceeding 7, indicative of higher levels of complexity.

\section{Dataset Construction}
\label{sec:pipeline}

\subsection{Overview}
\begin{figure}[t]
    \centering
    \includegraphics[width=1.0\linewidth]{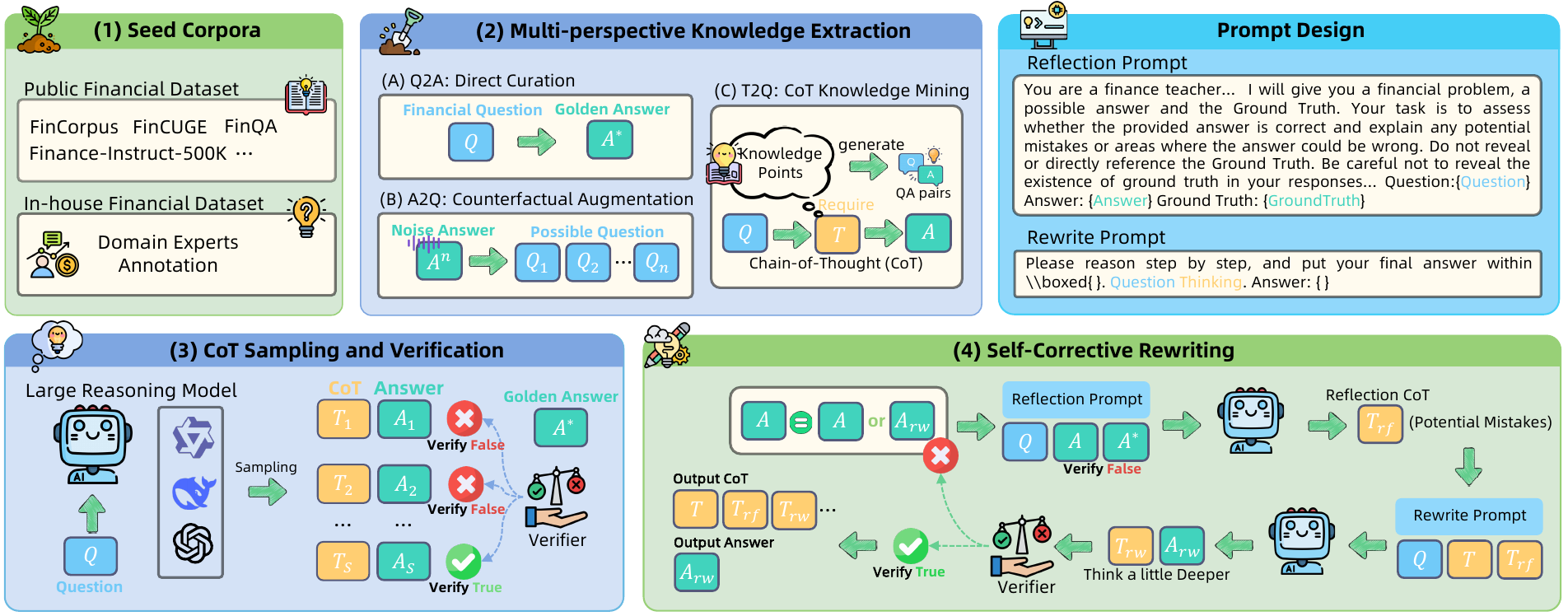}
    \caption{Illustration of the pipeline for constructing Agentar-DeepFinance-100K .}
    \vspace{-10pt}
    \label{fig:pipeline}
\end{figure}

Our proposed dataset construction pipeline is illustrated in Figure~\ref{fig:pipeline}. The pipeline consists of three main stages: (1) Starting from seed corpora, multi-perspective knowledge extraction is first adopted to generate diverse QA pairs. (2) Sample candidate CoTs from current advanced large reasoning models (LRMs)~\cite{guo2025deepseek,openaio1,qwen2.5} and perform answer consistency verification against golden answers. (3) Implement self-corrective rewriting on QA pairs that cannot be sampled correctly. Our comprehensive pipeline facilitates systematically producing high-quality, diverse and challenging reasoning trajectories.

\subsection{Seed Corpora}
The seed corpora serve as the foundational component for dataset construction, requiring extensive coverage across diverse financial domains. Existing public financial datasets~\cite{chen2021finqa,chen2022convfinqa,liu2025fin} provide structured knowledge from standardized materials (e.g., examinations), restricting their capacity to capture real-world user interactions. To bridge this gap, we systematically curated a large proprietary dataset with real domain expert annotations.

\subsection{Multi-perspective Knowledge Extraction}
The knowledge extraction phase aims to generate high-quality and rich QA pairs, which can model both explicit and implicit knowledge from the seed corpus. To achieve this, we propose the Multi-perspective Knowledge Extraction (MKE), comprising three extraction approaches:
\begin{itemize}
    \item \textbf{Q2A (Direct Curation)}: It serves as the foundation by directly harvesting well-structured QA pairs from the seed corpora. Subsequently, removing duplicates and filtering out low-quality content ensures the relevance and quality.

    \item \textbf{A2Q (Counterfactual Augmentation)}: The A2Q mechanism implements counterfactual exploration to enhance the knowledge space's density and causal connectivity. By perturbing original answers through semantic negation and contextual antonym substitution, adversarial answer variants are constructed. Subsequently, we adopt LRMs to generate questions for these answers, followed by a multi-stage verification protocol assessing semantic coherence, logical consistency and answer faithfulness. This bidirectional adversarial synthesis expands the neighborhood of core knowledge points.

    \item \textbf{T2Q (CoT Knowledge Mining)}: During the thinking phase of LRM, additional knowledge points may be introduced in CoTs. Sometimes, these knowledge points are not common sense but crucial to the reasoning process. Therefore, we developed a methodology to extract this latent knowledge from CoTs. The approach involves summarizing CoTs, prompting LRMs to identify key processes, and creating QA pairs. This process captures emergent knowledge points that bridge the gap between surface-level QA and deep reasoning requirements.
\end{itemize}
Our multi-perspective knowledge extraction framework achieves exhaustive and comprehensive knowledge modeling: Q2A preserves canonical patterns, A2Q strengthens causal connectivity through counterfactual perturbations, and T2Q uncovers implicit reasoning dependencies. The experiment in Section~\ref{sec:alb} demonstrates the effectiveness of this paradigm.

\subsection{CoT Sampling and Verification}
After multi-perspective knowledge extraction, numerous QA pairs are generated. To further construct high-quality reasoning traces, we first sample multiple CoTs and corresponding answers for each QA pair. Noticing that regular expression-based verification methods are insufficient for financial QA due to answer format variability (e.g., monetary expressions), we employ a lightweight model to match the answers. Since the verification task is simple and well-defined, a small model is sufficient and more cost-effective than a larger one. Only rigorously verified QA pairs with matching numerical precision and logical coherence are retained, ensuring the dataset's reliability for complex financial reasoning tasks.

\subsection{Self-Corrective Rewriting}
While modern LRMs exhibit impressive performance, significant challenges persist in addressing complex financial problems. A common practice in existing methodologies is to discard QA pairs that lack correctly sampled answers, thereby ensuring the overall accuracy of the dataset. However, these discarded QA pairs often involve intricate reasoning processes, which could serve as valuable resources for enhancing model capabilities. Therefore, we came up with this problem: \textit{Can flawed reasoning trajectories be transformed into effective learning opportunities through appropriate correction mechanisms?}

To accomplish this, we introduce Self-Corrective Rewriting (SCR) to enable the model to refine its answers by giving it some insights from the golden answers. More specifically, the SCR comprises two main steps: (1) \textbf{Reflection phase}: The model examines the disparities between its incorrect answer $A$ and the golden answer $A^{*}$ to generate diagnostic reflections CoT $T_{rf}$, which contains potential mistakes. (2) \textbf{Rewriting phase}: The reflection CoT $T_{rf}$ is merged with the original reasoning trace $T$. The model then continues the generation process based on this merged CoT to produce a new CoT $T_{rw}$ and the revised answer $A_{rw}$. Subsequently, $A_{rw}$ is verified against the correct answer $A^{*}$. If the verification is successful, exit the SCR process. The final CoT output consists of the original CoT, with the reflection CoT $T_{rf}$ and rewritten CoT $T_{rw}$ from each iteration alternately appended. The ultimate answer output is $A_{rw}$. If the verification fails, the correction cycle persists until the iteration limit is reached.

After SCR, more challenging questions can be correctly sampled. It is observed that the CoTs generated by SCR become much longer than before. Experiments in Section~\ref{sec:alb} have shown that integrating these challenging questions' CoTs obtained through SCR can significantly boost the model's reasoning capabilities.

\section{Experiment}
\label{sec:exp}

\noindent \textbf{Evaluation Setup.}
For rapid experiments, we randomly selected 60K samples from our rigorously quality-filtered dataset. The experiments were conducted on the ms-swift framework~\cite{zhao2024swift}, with the Qwen2.5-7B-Instruct model trained for 3 epochs. Model performance was assessed on three financial benchmarks: FinQA~\cite{chen2021finqa}, Fin-Eva~\cite{fineva} and a self-sampled in-domain subset. Among these, Fin-Eva consists of multiple-choice and true/false questions that can be verified through rule-based evaluation. In contrast, FinQA and our in-domain test set include open-ended questions with answers that are difficult to parse using regular expressions, necessitating the use of the LLM-as-a-judge evaluation approach. Fin-Eva's test set is divided into two parts: the Ant Group and the SUFE data sources. The results presented in this paper are based on data from the Ant Group source. Additionally, to assess general reasoning capabilities, we evaluated two widely-used benchmarks: MATH-500~\cite{hendrycks2021measuring} and GPQA-Diamond~\cite{rein2024gpqa}. The pass@1 accuracy is reported with $1$ and $8$ responses per query for MATH-500 and GPQA-Diamond, respectively.

\subsection{Investigation on CoT necessity}
\label{necessity}
Existing works typically synthesize CoTs exclusively for questions with formally verifiable answers, leaving it unclear which financial question types can benefit from CoTs. In this section, we conduct experiments identifying financial questions where CoT yields gains. Specifically, we explore two classification schemes: one based on task type and the other on question difficulty. Our experiments are performed on the rigorously filtered subset of 60K samples.

\begin{wraptable}{r}{0.46\textwidth}
\centering
\small
\resizebox{\linewidth}{!}{
\begin{tabular}{llcc}
\toprule[1pt]
                        \multicolumn{2}{l}{Method}                & FinQA & Fin-Eva \\ \midrule
               \multicolumn{2}{l}{Qwen2.5-7B-Instruct}   & 63.47 & 83.70   \\ \midrule
\multirow{2}{*}{Simple} & $w/o$ CoT           & 67.65 & 84.21   \\
                        & $w/\text{-}$ CoT            & \textbf{68.96} & \textbf{85.01}   \\ \midrule
\multirow{2}{*}{Hard}   & $w/o$ CoT           & 63.03 & 75.80   \\
                        & $w/\text{-}$ CoT            & \textbf{69.31} & \textbf{85.96}   \\ \bottomrule[1pt]
\end{tabular}
}
\caption{CoT necessity analysis on difficulty.}
\label{tab:diff_necessity}
\label{tab:sota}
\end{wraptable}

\noindent \textbf{Difficulty.}
To validate the relationship between question difficulty and the necessity of CoTs, we classify questions into simple and hard subsets based on correctness labels from superior LLMs (e.g., Qwen2.5-14B-Instruct). For a fair comparison, equal sample sizes were maintained for both subsets. Table~\ref{tab:diff_necessity} presents the results with and without CoTs on these two types of questions. It demonstrates that CoT yields gains on both simple and hard questions and improves more significantly on hard problems. Notably, when fine-tuning on hard questions without CoTs, a performance decline was observed. It suggests that complex financial reasoning may not be sufficiently learned by models lacking explicit CoT guidance.

\begin{table}[h]
\resizebox{\textwidth}{!}{
\begin{tabular}{l|cccccc|c}
\toprule[1pt]
CoT     & Text Gen.   & Analysis \& Interp.  & NLP    & Compliance \& Sec.    & Math       & Knowledge QA   & Average \\ \midrule
$w/o$      & 74.55       & 81.36                & 76.71  & 91.25                 & 31.71      & 78.32          & 72.32        \\
$w/\text{-}$       & \textbf{83.64}       & \textbf{88.05}                & \textbf{77.21}  & \textbf{92.02}                 & \textbf{61.15}      & \textbf{79.01}          &  \textbf{80.18}       \\ \bottomrule[1pt]
\end{tabular}
}
\caption{CoT necessity analysis on task type.}
\label{tab:necessity}
\end{table}

\noindent \textbf{Task Type.}
We partition the dataset into 6 subsets aligned with the financial task taxonomy defined in Section~\ref{sec:overview}. For each subset, experiments with and without CoT integration are conducted, reserving $10\%$ for testing. Additionally, the security compliance test set also includes the security compliance cases from Fin-Eva for more objective evaluation. As shown in Table~\ref{tab:necessity}, CoT integration consistently improves performance across all tasks. Notably, reasoning-intensive tasks, such as financial mathematics, exhibit substantial gains, demonstrating the critical role of CoT in complex financial reasoning.

\begin{AIbox}{Takeaway 4.1 for CoT Necessity}
Incorporating CoT consistently improves model performance regardless of task type and difficulty, particularly in complex reasoning tasks (e.g., math) and difficult problems.
\end{AIbox}

\subsection{Investigation on CoT Synthesizer}
\label{sec:cot_synthesizer}
With the rapid advancement of large reasoning models, the research community has witnessed the emergence of numerous models, characterized by enhanced general reasoning capabilities. Many existing studies~\cite{liu2025fin, zhu2025dianjin} predominantly employ DeepSeek-R1 for distilling financial CoTs, with limited investigation into CoT synthesizers. This lack of exploration leaves unanswered questions regarding the impact of different reasoning models on CoT synthesis and which model is best suited for synthesizing financial CoTs.

\begin{table}[htbp]
  \centering
  \resizebox{\textwidth}{!}{
  \begin{tabular}{@{}l c c c c c c c@{}}
    \toprule[1pt]
    \multirow{2}{*}{Model} & \multirow{2}{*}{Dense} & 
    \multicolumn{2}{c}{Fin-Eva} & \multicolumn{2}{c}{FinQA} & 
    \multicolumn{2}{c}{General} \\
    \cmidrule(lr){3-4} \cmidrule(lr){5-6} \cmidrule(lr){7-8}
    & & Acc & Res. Length & Acc & Res. Length & MATH & GPQA \\
    \midrule
    DeepSeek-R1       & \ding{55}     & \textbf{91.42}   & \textbf{770.14}   & \textbf{75.41}   & 675.46   & \textbf{98.20}    & 61.62   \\
    QwQ-Plus          & \ding{51}     & 90.99   & 627.43   & 72.97   & \textbf{1831.49}  & 97.20    & 54.55   \\
    Qwen3-235B-A22B   & \ding{55}     & 90.01   & 428.54   & 74.54   & 1302.86  & 94.60    & \textbf{62.12}   \\
    Qwen2.5-7B-Instruct & \ding{51}   & 83.70   & 1.00     & 63.47   & 239.97   &   73.20    & 37.88   \\
    \midrule[1pt]
    \midrule[1pt]
    \multirow{2}{*}{Teacher Model} & \multirow{2}{*}{\#Samples/Avg. Length} & 
    \multicolumn{2}{c}{Fin-Eva} & \multicolumn{2}{c}{FinQA} & 
    \multicolumn{2}{c}{General} \\
    \cmidrule(lr){3-4} \cmidrule(lr){5-6} \cmidrule(lr){7-8}
    & & Acc & Res. Length & Acc & Res. Length & MATH & GPQA \\
    \midrule
    DeepSeek-R1       & 21,040/1564.63  & 85.32   & \textbf{754.54}   & 66.52   & 1270.96  & 80.00     & 41.41     \\
    QwQ-Plus          & 19,323/1795.99  & \textbf{85.53}   & 702.41   & \textbf{68.79}   & \textbf{2196.81}  & \textbf{81.60}    & \textbf{42.42}     \\
    Qwen3-235B-A22B   & 20,322/1522.92  & 85.45   & 640.81   & 66.43   & 1469.85  & 80.00     & \textbf{42.42}     \\
    \bottomrule[1pt]
  \end{tabular}
}
\caption{Performance comparison of various CoT synthesizers, including their intrinsic performance and the performance of corresponding distilled student models. Additionally, the average response lengths on Fin-Eva and FinQA are reported.}
\label{tab:cot_synthesizer}
\end{table}

In this section, we employ multiple general reasoning models to synthesize CoTs for the same financial dataset, aiming to identify the most effective synthesizer. Specifically, a subset of 25K samples was randomly selected, with each model generating four CoTs and corresponding answers per question. Only questions containing the correct sampled answer were retained for analysis. Table~\ref{tab:cot_synthesizer} summarizes the experimental results, comparing both the intrinsic reasoning capabilities of the models and their effectiveness in distilling knowledge to improve student models. It reveals a key insight: \textit{a good model is not necessarily good at teaching students.}

It shows that DeepSeek-R1 demonstrated superior performance not only in financial reasoning tasks but also in general knowledge evaluation, compared with QwQ-Plus and Qwen3-235B-A22B. Moreover, DeepSeek-R1 sampled more questions correctly during the process of sampling CoTs for the 25K samples. It can also be observed that DeepSeek-R1 produced CoTs in shorter and more stable lengths when answering questions in Fin-Eva and FinQA. However, the performance of smaller models fine-tuned using CoTs generated by DeepSeek-R1 was suboptimal. Interestingly, QwQ-plus, a model with relatively weaker financial and general capabilities, achieved the best overall performance. Specifically, compared with DeepSeek-R1, the model distilled from QwQ-Plus achieved improvements of +0.2 and +2.2 on Fin-Eva and FinQA, respectively, as well as increases of +1.6 and +1.0 on MATH and GPQA. Further analysis revealed that CoTs generated by QwQ-Plus tend to be longer than those synthesized by DeepSeek-R1. This characteristic likely contributed to the superior performance of models fine-tuned with QwQ-Plus-generated CoTs. Consequently, we adopt QwQ as the primary generator for financial CoT synthesis.

\begin{AIbox}{Takeaway 4.2 for CoT Synthesizer}
The effectiveness of a reasoning model as a CoT synthesizer does not always align with its intrinsic reasoning performance.
\end{AIbox}

\subsection{Investigation on CoT Length}
To investigate the impact of CoT length, we employ a prompt-based approach to generate CoTs of varying lengths from the same CoT synthesizer. Specifically, except for the original CoTs, which are recognized as long CoTs, we generate two additional shorter versions of CoTs: (1) Middle-length CoTs: Inspired by CCoT~\cite{renze2024benefits}, we append the phrase "Be concise" to the instruction, which significantly reduces the CoT length. (2) Short-length CoTs:  We select samples where the model exhibits overthinking and manually simplify them. The model then generates shorter CoTs via in-context learning using these simplified samples.

Based on the findings in Section~\ref{sec:cot_synthesizer}, the CoT synthesizer employed in this section is QwQ-Plus. The experimental results presented in Table~\ref{tab:qwq_cot_length_comparison} validate the effectiveness of the prompt-based approach in controlling CoT length while maintaining overall performance. In terms of CoT length, the generated responses on FinQA were reduced by $25.0\%$ and $29.6\%$ compared to the long CoTs. The only notable difference was that using short CoTs led to a $1.5$ decrease in accuracy on Fin-Eva.

\begin{table}[h]
\centering
\begin{tabular}{@{}l c c c c@{}}
\toprule[1pt]
\multirow{2}{*}{CoT Length} & \multicolumn{2}{c}{Fin-Eva} & \multicolumn{2}{c}{FinQA} \\
\cmidrule(lr){2-3} \cmidrule(lr){4-5}
                             & Acc       & Res. Length     & Acc       & Res. Length    \\
\midrule
Long                         & 90.99     & \textbf{627.43}          & \textbf{72.97}     & \textbf{1831.49}        \\
Medium                       & \textbf{91.07}     & 562.98          & 72.45     & 1374.59        \\
Short                        & 89.51     & 459.48          & 72.97     & 1288.94        \\
\bottomrule[1pt]
\end{tabular}
\caption{Performance comparison of QwQ-Plus under different CoT lengths on Fin-Eva and FinQA. The medium- and short-length CoTs are generated via our prompt-based CoT length control method.}
\label{tab:qwq_cot_length_comparison}
\end{table}

\begin{table}[h]
\centering
\begin{tabular}{@{}l c c c c c@{}}
\toprule[1pt]
\multirow{2}{*}{CoT Length} & \multirow{2}{*}{Avg. Length} & \multicolumn{2}{c}{Fin-Eva} & \multicolumn{2}{c}{FinQA} \\
\cmidrule(lr){3-4} \cmidrule(lr){5-6}
                             &                                  & Acc       & Res. Length     & Acc       & Res. Length     \\
\midrule
Long                         & 1352.16                          & \textbf{85.53}     & 702.41          & \textbf{68.79}    & \textbf{2196.81}        \\
Medium                       & 1248.42                          & 85.40     & \textbf{742.90}  & 65.13      & 1698.63          \\
Short                        & 1020.26                          & 85.19     & 574.00          & 65.56      & 1587.76          \\
\bottomrule[1pt]
\end{tabular}
\caption{Comparison of the models distilled from QwQ-Plus with different CoT lengths. The average CoT lengths in the training corpora are reported.}
\label{tab:cot_length_distill}
\end{table}

Table~\ref{tab:cot_length_distill} illustrates the performance of the model distilled using CoTs of three different lengths. It further confirms the effectiveness of our prompt-based CoT length control method, with medium-length and short-length CoTs resulting in reductions of $7.7\%$ and $24.5\%$, respectively. Notably, the model trained with the longest CoT demonstrates superior performance on both the Fin-Eva and FinQA benchmarks. Specifically, it achieves improvements of +$3.7$ and +$3.2$ on FinQA compared to models trained with medium-length and short CoTs, respectively. In conclusion, while reducing the CoT length in the training corpus can result in more concise model responses, it may also compromise the model's overall performance.

\begin{AIbox}{Takeaway 4.3 for CoT Length}
Distilling CoTs with reduced length allows the trained model to provide more concise responses, but may also hurt performance. Financial reasoning requires long CoTs.
\end{AIbox}

\subsection{Ablation Study}
\label{sec:alb}

\begin{figure*}[htbp]
\centering
\begin{minipage}{0.498\columnwidth}
\centering
\includegraphics[width=\linewidth]{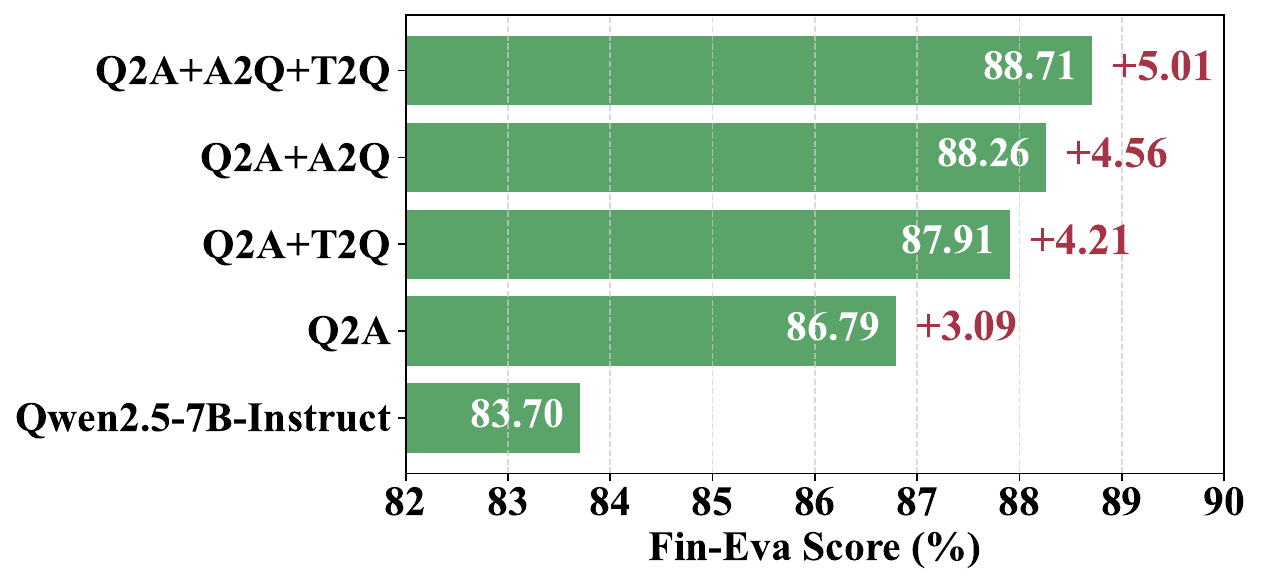}
\caption{Ablation Study for Multi-perspective Knowledge Extraction.}
\label{fig:mke}
\end{minipage}
\hfill
\begin{minipage}{0.48\columnwidth}
\centering
\includegraphics[width=\linewidth]{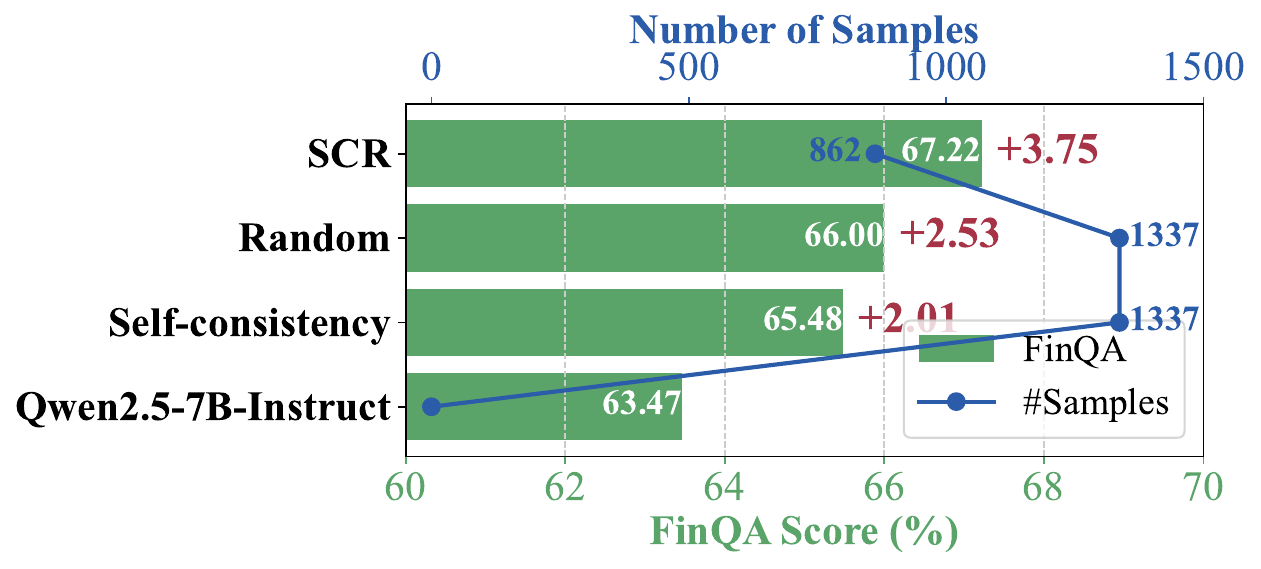}
\caption{Ablation Study for Self-Corrective Rewriting.}
\label{fig:scr}
\end{minipage}
\end{figure*}

\noindent \textbf{Impact of Multi-perspective Knowledge Extraction.}
To evaluate the effectiveness of our proposed Multi-perspective Knowledge Extraction (MKE), we conducted experiments with the 60K samples subset (denoted as Q2A). In this process, we employed the A2Q and T2Q synthesis methods within MKE to create two new datasets: one with 19K A2Q samples and another with 34K T2Q samples. The experimental results are shown in Figure~\ref{fig:mke}. It shows that training with only the Q2A dataset yields a $+3.09$ improvement on Fin-Eva. Moreover, incorporating the T2Q dataset with Q2A enhances performance by $+1.12$, and incorporating A2Q with Q2A yields $+1.47$. Furthermore, using the A2Q and T2Q datasets for joint training yields a significant improvement of +$1.92$.

\noindent \textbf{Impact of Self-Corrective Rewriting.}
To evaluate the effectiveness of SCR, we conducted experiments using the FinQA training set (5K samples) and reported the results on the FinQA test set. We sampled 4 candidate answers for each question through QwQ-Plus and found that 1,337 samples lacked correct answers after verification. Figure~\ref{fig:scr} presents our experimental results. For these 1,337 incorrectly sampled cases, we compared training performance using self-consistency and random sampling. The results show improved performance when training on incorrectly sampled data, with random sampling outperforming self-consistency. In addition, after training with 862 samples containing correct answers generated after applying SCR, the performance continued to improve by $+1.74$, validating the efficacy of SCR.

\section{Related Work}
\label{sec:related}

\noindent \textbf{Reasoning Model.}
The initial developments in reasoning models stemmed from approaches involving few-shot prompting with decomposed reasoning steps~\cite{wei2022chain} and zero-shot activation through trigger phrases like "Let's think step by step"~\cite{kojima2022large}. Progress in this field was further driven by RL-based train-time scaling~\cite{zhang2024rest} and search-based test-time scaling~\cite{snell2024scaling}, indicating a promising direction for enhancing the reasoning capacities of LLMs. A significant milestone in this direction is OpenAI o1 series~\cite{openaio1}, which showcased impressive reasoning capabilities. Subsequently, a series of open-source o1 reproductions emerged, such as OpenR~\cite{wang2024openr}, LLaVA-o1~\cite{xu2024llava}, o1-Coder~\cite{zhang2024o1}, LLaMA-Berry~\cite{zhang2024llama} and Journey Learning~\cite{qin2024o1,huang2024o1,huang2025o1replicationjourney}. Recently, DeepSeek-R1~\cite{guo2025deepseek} introduced long CoT cold start and rule-based RL, proving that simply verifying the final answer can allow the model to automatically explore and generate long CoTs, thereby improving its reasoning ability. Subsequent research endeavors have focused on replicating R1 (e.g., OpenR1~\cite{openr1}) or applying it to various domains, such as R1-Omni~\cite{zhao2025r1} in sentiment analysis, MedVLM-R1~\cite{pan2025medvlm} in medical imaging and Fin-R1~\cite{liu2025fin} in finance.

\noindent \textbf{Financial LLMs.}
The Finance LLM is designed to enhance the general LLM's capabilities in the financial domain. Early efforts, such as FinBERT~\cite{araci2019finbert}, built on BERT and focused on tasks like sentiment analysis and financial text classification. With the success of GPT models, domain-specific models such as BloombergGPT~\cite{wu2023bloomberggpt}, FinMA~\cite{xie2023pixiu}, FinGPT~\cite{wang2023fingpt}, InvestLM~\cite{yang2023investlm} have emerged, leveraging instruction tuning and curated datasets to improve performance on financial tasks. Recent advancements have integrated reasoning capability, as seen in O1-based models such as XuanYuan-FinX1-Preview~\cite{XuanYuan2024} and Fino1~\cite{qian2025fino1}, which demonstrate enhanced proficiency in handling complex financial reasoning tasks. Furthermore, Fin-R1~\cite{liu2025fin} and DianJin-R1~\cite{zhu2025dianjin} distill cold-start CoTs from DeepSeek-R1~\cite{guo2025deepseek} and perform rule-based RL on hard-to-sample examples.

\noindent \textbf{CoT Synthesis.}
Initially, the CoTs generated by the few-shot~\cite{wei2022chain} and zero-shot~\cite{kojima2022large} methods are short. Moreover, the CoT examples adopted in few-shot methods are manually designed. Subsequent advancements~\cite{qi2024mutual,guan2025rstar} integrated reward models (e.g., PRM~\cite{lightman2023let}) to guide the search of high-quality CoTs. More recently, with the emergence of powerful reasoning models~\cite{guo2025deepseek,openaio1,qwen2.5}, distillation~\cite{he2025deepmath,muennighoff2025s1,openr1} from these models emerged as a cost-effective strategy for CoT Synthesis. This process involves querying large language models (LLMs) to generate CoTs and answers, followed by verification to ensure consistency. Additionally, post-processing techniques, such as token compression for efficiency~\cite{xia2025tokenskip} and length extension for richer reasoning~\cite{shen2025long}, have further refined the distilled CoTs.

\section{Conclusion}
\label{sec:conclusion}
This paper presents Agentar-DeepFinance-100K , a large-scale financial reasoning dataset constructed with a systematically optimized chain-of-thought (CoT) synthesis framework. We propose a pipeline incorporating multi-perspective knowledge extraction and self-corrective rewriting to produce comprehensive reasoning trajectories. Additionally, we systematically investigate critical factors influencing CoT effectiveness, providing empirical insights for high-quality financial CoT construction. Experiments demonstrate the effectiveness of our Agentar-DeepFinance-100K , with significant improvements on financial reasoning benchmarks of models trained on our dataset. Future work will focus on exploring multi-modal financial reasoning, incorporating visual financial content such as charts.

\clearpage
\bibliographystyle{unsrtnat} 
\bibliography{main}

\newpage
\section*{Appendix}

\begin{figure*}[htbp]
    \centering
    \includegraphics[width=0.95\linewidth]{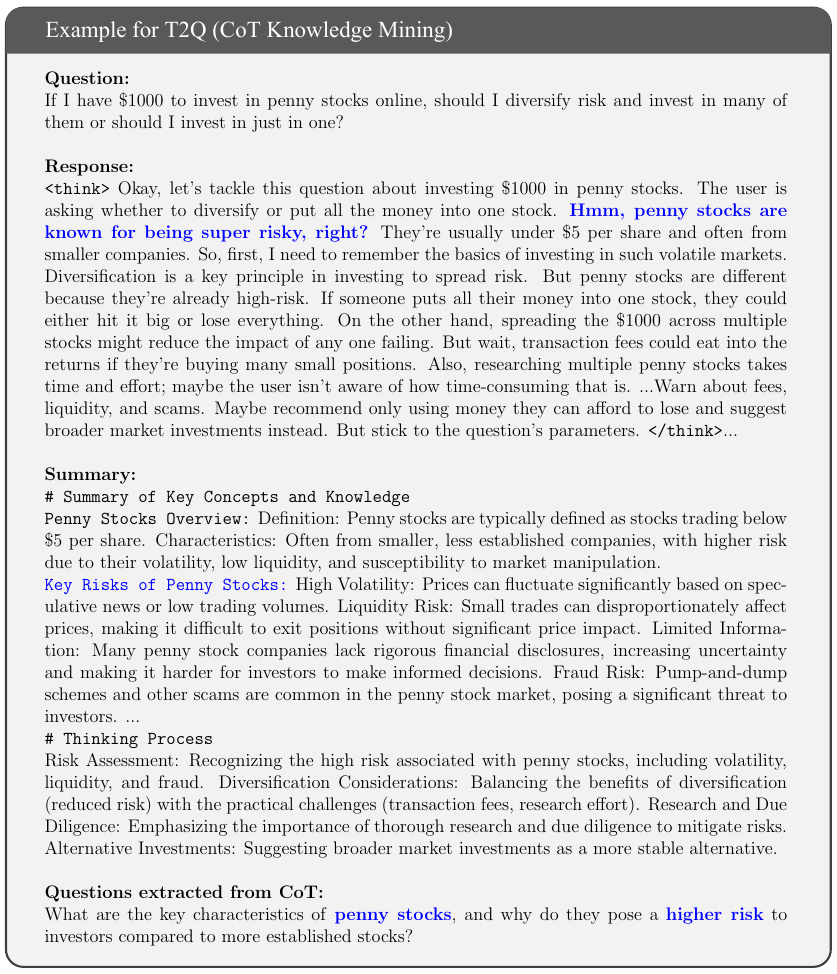}
    \caption{An example for T2Q (CoT knowledge) in multi-perspective knowledge extraction.}
    \label{fig:cot_extraction}
\end{figure*}

\begin{figure*}[htbp]
    \centering
    \includegraphics[width=0.95\linewidth]{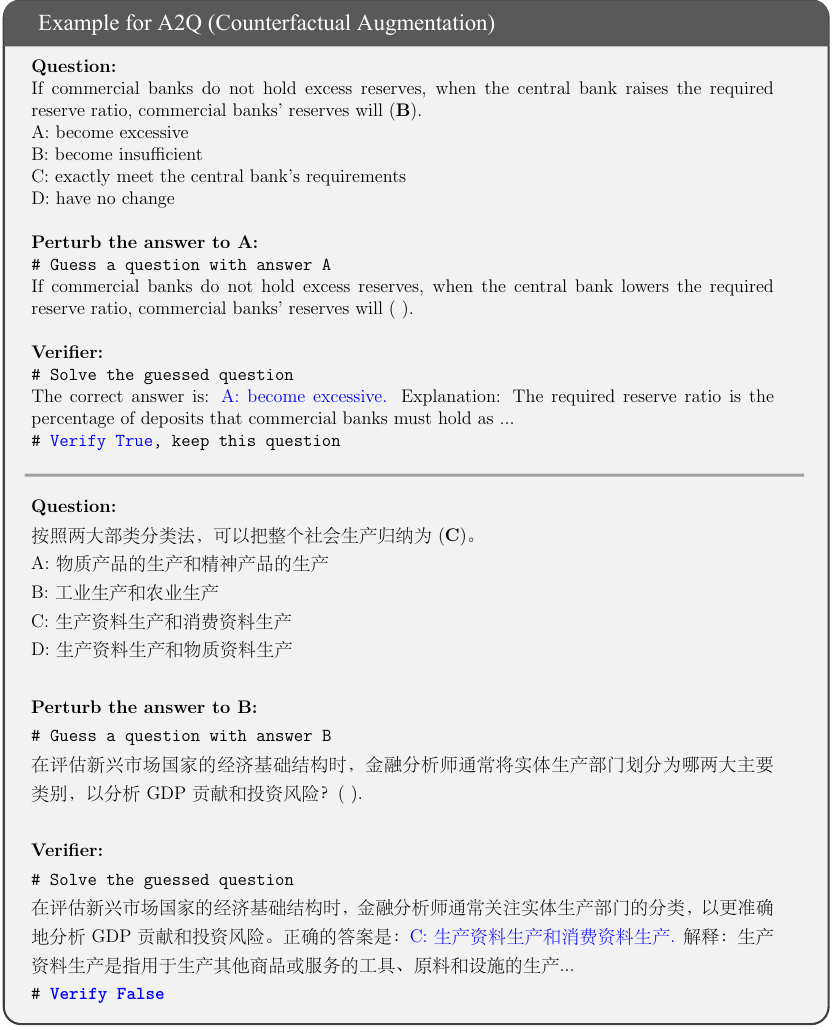}
    \caption{An example for A2Q (Counterfactual Augmentation) in multi-perspective knowledge extraction.}
    \label{fig:a2q_example}
\end{figure*}

\begin{figure*}[htbp]
    \centering
    \includegraphics[width=0.95\linewidth]{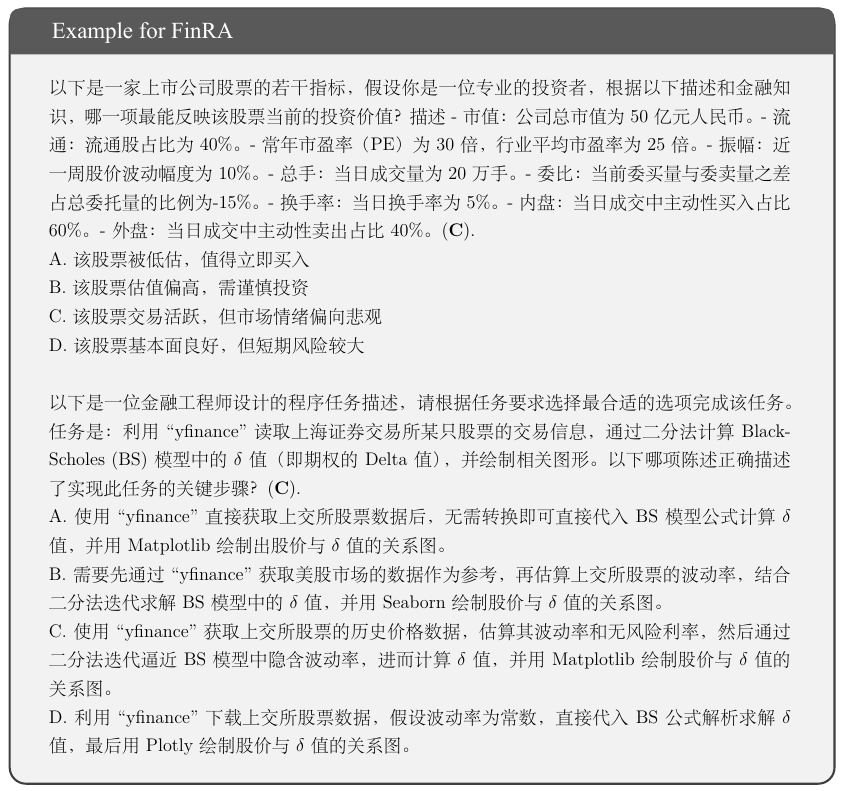}
    \caption{Examples for our in-house financial dataset FinRA.}
    \label{fig:finra_example}
\end{figure*}

\end{document}

%% file: commands.tex
\renewcommand{\phi}{\varphi}

\renewcommand{\epsilon}{\varepsilon}
\renewcommand{\imath}{\mathrm{i}}

\newlength{\restsubwidth}
\newlength{\restsubheight}
\newlength{\restsubmoreheight}
\newcommand{\rest}[2]{%
        \settowidth{\restsubwidth}{\ensuremath{#2}}
        \settoheight{\restsubheight}{\ensuremath{{}_{#2}}}
        \ensuremath{{#1\hskip 0.5pt}_{\vrule\kern2pt\parbox[b][%
        4pt][b]{\the\restsubwidth}{%
                        \ensuremath{{}_{#2}}}}}
        }

%% file: main.bbl
\begin{thebibliography}{47}
\providecommand{\natexlab}[1]{#1}
\providecommand{\url}[1]{\texttt{#1}}
\expandafter\ifx\csname urlstyle\endcsname\relax
  \providecommand{\doi}[1]{doi: #1}\else
  \providecommand{\doi}{doi: \begingroup \urlstyle{rm}\Url}\fi

\bibitem[Xie et~al.(2023)Xie, Han, Zhang, Lai, Peng, Lopez-Lira, and Huang]{xie2023pixiu}
Qianqian Xie, Weiguang Han, Xiao Zhang, Yanzhao Lai, Min Peng, Alejandro Lopez-Lira, and Jimin Huang.
\newblock Pixiu: A large language model, instruction data and evaluation benchmark for finance.
\newblock \emph{arXiv preprint arXiv:2306.05443}, 2023.

\bibitem[Wang et~al.(2023)Wang, Yang, and Wang]{wang2023fingpt}
Neng Wang, Hongyang Yang, and Christina~Dan Wang.
\newblock Fingpt: Instruction tuning benchmark for open-source large language models in financial datasets.
\newblock \emph{arXiv preprint arXiv:2310.04793}, 2023.

\bibitem[Yang et~al.(2023)Yang, Tang, and Tam]{yang2023investlm}
Yi~Yang, Yixuan Tang, and Kar~Yan Tam.
\newblock Investlm: A large language model for investment using financial domain instruction tuning.
\newblock \emph{arXiv preprint arXiv:2309.13064}, 2023.

\bibitem[Araci(2019)]{araci2019finbert}
Dogu Araci.
\newblock Finbert: Financial sentiment analysis with pre-trained language models.
\newblock \emph{arXiv preprint arXiv:1908.10063}, 2019.

\bibitem[Yu et~al.(2023)Yu, Chen, Ling, Dong, Liu, and Lu]{yu2023temporal}
Xinli Yu, Zheng Chen, Yuan Ling, Shujing Dong, Zongyi Liu, and Yanbin Lu.
\newblock Temporal data meets llm--explainable financial time series forecasting.
\newblock \emph{arXiv preprint arXiv:2306.11025}, 2023.

\bibitem[Kou et~al.(2024)Kou, Yu, Luo, Peng, and Chen]{kou2024automate}
Zhizhuo Kou, Holam Yu, Junyu Luo, Jingshu Peng, and Lei Chen.
\newblock Automate strategy finding with llm in quant investment.
\newblock \emph{arXiv preprint arXiv:2409.06289}, 2024.

\bibitem[OpenAI(2024)]{openaio1}
OpenAI.
\newblock Openai o1.
\newblock \url{https://openai.com/o1}, 2024.
\newblock Accessed: 2025-03-12.

\bibitem[Guo et~al.(2025)Guo, Yang, Zhang, Song, Zhang, Xu, Zhu, Ma, Wang, Bi, et~al.]{guo2025deepseek}
Daya Guo, Dejian Yang, Haowei Zhang, Junxiao Song, Ruoyu Zhang, Runxin Xu, Qihao Zhu, Shirong Ma, Peiyi Wang, Xiao Bi, et~al.
\newblock Deepseek-r1: Incentivizing reasoning capability in llms via reinforcement learning.
\newblock \emph{arXiv preprint arXiv:2501.12948}, 2025.

\bibitem[Yang et~al.(2024)Yang, Yang, Zhang, Hui, Zheng, Yu, Li, Liu, Huang, Wei, Lin, Yang, Tu, Zhang, Yang, Yang, Zhou, Lin, Dang, Lu, Bao, Yang, Yu, Li, Xue, Zhang, Zhu, Men, Lin, Li, Tang, Xia, Ren, Ren, Fan, Su, Zhang, Wan, Liu, Cui, Zhang, and Qiu]{qwen2.5}
An~Yang, Baosong Yang, Beichen Zhang, Binyuan Hui, Bo~Zheng, Bowen Yu, Chengyuan Li, Dayiheng Liu, Fei Huang, Haoran Wei, Huan Lin, Jian Yang, Jianhong Tu, Jianwei Zhang, Jianxin Yang, Jiaxi Yang, Jingren Zhou, Junyang Lin, Kai Dang, Keming Lu, Keqin Bao, Kexin Yang, Le~Yu, Mei Li, Mingfeng Xue, Pei Zhang, Qin Zhu, Rui Men, Runji Lin, Tianhao Li, Tianyi Tang, Tingyu Xia, Xingzhang Ren, Xuancheng Ren, Yang Fan, Yang Su, Yichang Zhang, Yu~Wan, Yuqiong Liu, Zeyu Cui, Zhenru Zhang, and Zihan Qiu.
\newblock Qwen2.5 technical report.
\newblock \emph{arXiv preprint arXiv:2412.15115}, 2024.

\bibitem[Liu et~al.(2025)Liu, Guo, Lou, Zeng, Niu, Wang, Xu, Cai, Yang, Zhao, et~al.]{liu2025fin}
Zhaowei Liu, Xin Guo, Fangqi Lou, Lingfeng Zeng, Jinyi Niu, Zixuan Wang, Jiajie Xu, Weige Cai, Ziwei Yang, Xueqian Zhao, et~al.
\newblock Fin-r1: A large language model for financial reasoning through reinforcement learning.
\newblock \emph{arXiv preprint arXiv:2503.16252}, 2025.

\bibitem[Zhu et~al.(2025)Zhu, Chen, Dou, Li, Guo, Chen, and Zhang]{zhu2025dianjin}
Jie Zhu, Qian Chen, Huaixia Dou, Junhui Li, Lifan Guo, Feng Chen, and Chi Zhang.
\newblock Dianjin-r1: Evaluating and enhancing financial reasoning in large language models.
\newblock \emph{arXiv preprint arXiv:2504.15716}, 2025.

\bibitem[Team(n.d.)]{FinCorpus}
Duxiaoman~DI Team.
\newblock Fincorpus.
\newblock \url{https://huggingface.co/datasets/Duxiaoman-DI/FinCorpus}, n.d.
\newblock Accessed: 2025-07-08.

\bibitem[Flowers(2025)]{josephgflowers2025financeinstruct}
Joseph~G. Flowers.
\newblock Finance-instruct-500k.
\newblock https://huggingface.co/datasets/Josephgflowers/Finance-Instruct-500k, 2025.
\newblock Accessed: 2025-07-08.

\bibitem[Lu et~al.(2023)Lu, Wu, Liang, Xu, He, Geng, Han, Xin, and Xiao]{lu2023bbt}
Dakuan Lu, Hengkui Wu, Jiaqing Liang, Yipei Xu, Qianyu He, Yipeng Geng, Mengkun Han, Yingsi Xin, and Yanghua Xiao.
\newblock Bbt-fin: Comprehensive construction of chinese financial domain pre-trained language model, corpus and benchmark.
\newblock \emph{arXiv preprint arXiv:2302.09432}, 2023.

\bibitem[Chen et~al.(2021)Chen, Chen, Smiley, Shah, Borova, Langdon, Moussa, Beane, Huang, Routledge, et~al.]{chen2021finqa}
Zhiyu Chen, Wenhu Chen, Charese Smiley, Sameena Shah, Iana Borova, Dylan Langdon, Reema Moussa, Matt Beane, Ting-Hao Huang, Bryan Routledge, et~al.
\newblock Finqa: A dataset of numerical reasoning over financial data.
\newblock \emph{arXiv preprint arXiv:2109.00122}, 2021.

\bibitem[csujeong(n.d.)]{FinancialData}
csujeong.
\newblock Financialdata.
\newblock \url{https://huggingface.co/datasets/csujeong/financial_data}, n.d.
\newblock Accessed: 2025-07-08.

\bibitem[Malik(n.d.)]{Quant-Trading-Instruct}
Lukas Malik.
\newblock Quant-trading-instruct.
\newblock \url{https://huggingface.co/datasets/lumalik/Quant-Trading-Instruct}, n.d.
\newblock Accessed: 2025-07-08.

\bibitem[Chen et~al.(2022)Chen, Li, Smiley, Ma, Shah, and Wang]{chen2022convfinqa}
Zhiyu Chen, Shiyang Li, Charese Smiley, Zhiqiang Ma, Sameena Shah, and William~Yang Wang.
\newblock Convfinqa: Exploring the chain of numerical reasoning in conversational finance question answering.
\newblock \emph{arXiv preprint arXiv:2210.03849}, 2022.

\bibitem[Zhao et~al.(2024)Zhao, Huang, Hu, Wang, Mao, Zhang, Jiang, Wu, Ai, Wang, Zhou, and Chen]{zhao2024swift}
Yuze Zhao, Jintao Huang, Jinghan Hu, Xingjun Wang, Yunlin Mao, Daoze Zhang, Zeyinzi Jiang, Zhikai Wu, Baole Ai, Ang Wang, Wenmeng Zhou, and Yingda Chen.
\newblock Swift:a scalable lightweight infrastructure for fine-tuning, 2024.
\newblock URL \url{https://arxiv.org/abs/2408.05517}.

\bibitem[Team(2024{\natexlab{a}})]{fineva}
Alipay Team.
\newblock Financial evaluation dataset.
\newblock \url{https://github.com/alipay/financial_evaluation_dataset}, 2024{\natexlab{a}}.
\newblock Accessed: 2025-03-18.

\bibitem[Hendrycks et~al.(2021)Hendrycks, Burns, Kadavath, Arora, Basart, Tang, Song, and Steinhardt]{hendrycks2021measuring}
Dan Hendrycks, Collin Burns, Saurav Kadavath, Akul Arora, Steven Basart, Eric Tang, Dawn Song, and Jacob Steinhardt.
\newblock Measuring mathematical problem solving with the math dataset.
\newblock \emph{arXiv preprint arXiv:2103.03874}, 2021.

\bibitem[Rein et~al.(2024)Rein, Hou, Stickland, Petty, Pang, Dirani, Michael, and Bowman]{rein2024gpqa}
David Rein, Betty~Li Hou, Asa~Cooper Stickland, Jackson Petty, Richard~Yuanzhe Pang, Julien Dirani, Julian Michael, and Samuel~R Bowman.
\newblock Gpqa: A graduate-level google-proof q\&a benchmark.
\newblock In \emph{First Conference on Language Modeling}, 2024.

\bibitem[Renze and Guven(2024)]{renze2024benefits}
Matthew Renze and Erhan Guven.
\newblock The benefits of a concise chain of thought on problem-solving in large language models.
\newblock In \emph{2024 2nd International Conference on Foundation and Large Language Models (FLLM)}, pages 476--483. IEEE, 2024.

\bibitem[Wei et~al.(2022)Wei, Wang, Schuurmans, Bosma, Xia, Chi, Le, Zhou, et~al.]{wei2022chain}
Jason Wei, Xuezhi Wang, Dale Schuurmans, Maarten Bosma, Fei Xia, Ed~Chi, Quoc~V Le, Denny Zhou, et~al.
\newblock Chain-of-thought prompting elicits reasoning in large language models.
\newblock \emph{Advances in neural information processing systems}, 35:\penalty0 24824--24837, 2022.

\bibitem[Kojima et~al.(2022)Kojima, Gu, Reid, Matsuo, and Iwasawa]{kojima2022large}
Takeshi Kojima, Shixiang~Shane Gu, Machel Reid, Yutaka Matsuo, and Yusuke Iwasawa.
\newblock Large language models are zero-shot reasoners.
\newblock \emph{Advances in neural information processing systems}, 35:\penalty0 22199--22213, 2022.

\bibitem[Zhang et~al.(2024{\natexlab{a}})Zhang, Zhoubian, Hu, Yue, Dong, and Tang]{zhang2024rest}
Dan Zhang, Sining Zhoubian, Ziniu Hu, Yisong Yue, Yuxiao Dong, and Jie Tang.
\newblock Rest-mcts*: Llm self-training via process reward guided tree search.
\newblock \emph{Advances in Neural Information Processing Systems}, 37:\penalty0 64735--64772, 2024{\natexlab{a}}.

\bibitem[Snell et~al.(2024)Snell, Lee, Xu, and Kumar]{snell2024scaling}
Charlie Snell, Jaehoon Lee, Kelvin Xu, and Aviral Kumar.
\newblock Scaling llm test-time compute optimally can be more effective than scaling model parameters.
\newblock \emph{arXiv preprint arXiv:2408.03314}, 2024.

\bibitem[Wang et~al.(2024)Wang, Fang, Wan, Wen, Zhu, Liu, Gong, Song, Chen, Ni, et~al.]{wang2024openr}
Jun Wang, Meng Fang, Ziyu Wan, Muning Wen, Jiachen Zhu, Anjie Liu, Ziqin Gong, Yan Song, Lei Chen, Lionel~M Ni, et~al.
\newblock Openr: An open source framework for advanced reasoning with large language models.
\newblock \emph{arXiv preprint arXiv:2410.09671}, 2024.

\bibitem[Xu et~al.(2024)Xu, Jin, Hao, Song, Sun, and Yuan]{xu2024llava}
Guowei Xu, Peng Jin, Li~Hao, Yibing Song, Lichao Sun, and Li~Yuan.
\newblock Llava-o1: Let vision language models reason step-by-step.
\newblock \emph{arXiv preprint arXiv:2411.10440}, 2024.

\bibitem[Zhang et~al.(2024{\natexlab{b}})Zhang, Wu, Yang, Shu, Xiao, Kong, and Sang]{zhang2024o1}
Yuxiang Zhang, Shangxi Wu, Yuqi Yang, Jiangming Shu, Jinlin Xiao, Chao Kong, and Jitao Sang.
\newblock o1-coder: an o1 replication for coding.
\newblock \emph{arXiv preprint arXiv:2412.00154}, 2024{\natexlab{b}}.

\bibitem[Zhang et~al.(2024{\natexlab{c}})Zhang, Wu, Lei, Che, Li, Xie, Huang, Zhang, Pavone, Li, et~al.]{zhang2024llama}
Di~Zhang, Jianbo Wu, Jingdi Lei, Tong Che, Jiatong Li, Tong Xie, Xiaoshui Huang, Shufei Zhang, Marco Pavone, Yuqiang Li, et~al.
\newblock Llama-berry: Pairwise optimization for o1-like olympiad-level mathematical reasoning.
\newblock \emph{arXiv preprint arXiv:2410.02884}, 2024{\natexlab{c}}.

\bibitem[Qin et~al.(2024)Qin, Li, Zou, Liu, Xia, Huang, Ye, Yuan, Liu, Li, et~al.]{qin2024o1}
Yiwei Qin, Xuefeng Li, Haoyang Zou, Yixiu Liu, Shijie Xia, Zhen Huang, Yixin Ye, Weizhe Yuan, Hector Liu, Yuanzhi Li, et~al.
\newblock O1 replication journey: A strategic progress report--part 1.
\newblock \emph{arXiv preprint arXiv:2410.18982}, 2024.

\bibitem[Huang et~al.(2024)Huang, Zou, Li, Liu, Zheng, Chern, Xia, Qin, Yuan, and Liu]{huang2024o1}
Zhen Huang, Haoyang Zou, Xuefeng Li, Yixiu Liu, Yuxiang Zheng, Ethan Chern, Shijie Xia, Yiwei Qin, Weizhe Yuan, and Pengfei Liu.
\newblock O1 replication journey--part 2: Surpassing o1-preview through simple distillation, big progress or bitter lesson?
\newblock \emph{arXiv preprint arXiv:2411.16489}, 2024.

\bibitem[Huang et~al.(2025)Huang, Geng, Hua, Huang, Zou, Zhang, Liu, and Zhang]{huang2025o1replicationjourney}
Zhongzhen Huang, Gui Geng, Shengyi Hua, Zhen Huang, Haoyang Zou, Shaoting Zhang, Pengfei Liu, and Xiaofan Zhang.
\newblock O1 replication journey -- part 3: Inference-time scaling for medical reasoning.
\newblock \emph{arXiv preprint arXiv:2501.06458}, 2025.

\bibitem[Face(2025)]{openr1}
Hugging Face.
\newblock Open r1: A fully open reproduction of deepseek-r1, January 2025.
\newblock URL \url{https://github.com/huggingface/open-r1}.

\bibitem[Zhao et~al.(2025)Zhao, Wei, and Bo]{zhao2025r1}
Jiaxing Zhao, Xihan Wei, and Liefeng Bo.
\newblock R1-omni: Explainable omni-multimodal emotion recognition with reinforcement learning.
\newblock \emph{arXiv preprint arXiv:2503.05379}, 2025.

\bibitem[Pan et~al.(2025)Pan, Liu, Wu, Liu, Zhu, Li, Chen, Ouyang, and Rueckert]{pan2025medvlm}
Jiazhen Pan, Che Liu, Junde Wu, Fenglin Liu, Jiayuan Zhu, Hongwei~Bran Li, Chen Chen, Cheng Ouyang, and Daniel Rueckert.
\newblock Medvlm-r1: Incentivizing medical reasoning capability of vision-language models (vlms) via reinforcement learning.
\newblock \emph{arXiv preprint arXiv:2502.19634}, 2025.

\bibitem[Wu et~al.(2023)Wu, Irsoy, Lu, Dabravolski, Dredze, Gehrmann, Kambadur, Rosenberg, and Mann]{wu2023bloomberggpt}
Shijie Wu, Ozan Irsoy, Steven Lu, Vadim Dabravolski, Mark Dredze, Sebastian Gehrmann, Prabhanjan Kambadur, David Rosenberg, and Gideon Mann.
\newblock Bloomberggpt: A large language model for finance.
\newblock \emph{arXiv preprint arXiv:2303.17564}, 2023.

\bibitem[Team(2024{\natexlab{b}})]{XuanYuan2024}
Duxiaoman~DI Team.
\newblock Xuanyuan-finx1-preview.
\newblock \url{https://github.com/Duxiaoman-DI/XuanYuan}, 2024{\natexlab{b}}.
\newblock Accessed: 2025-03-18.

\bibitem[Qian et~al.(2025)Qian, Zhou, Wang, Peng, Yi, Huang, Xie, and Nie]{qian2025fino1}
Lingfei Qian, Weipeng Zhou, Yan Wang, Xueqing Peng, Han Yi, Jimin Huang, Qianqian Xie, and Jianyun Nie.
\newblock Fino1: On the transferability of reasoning enhanced llms to finance.
\newblock \emph{arXiv preprint arXiv:2502.08127}, 2025.

\bibitem[Qi et~al.(2024)Qi, Ma, Xu, Zhang, Yang, and Yang]{qi2024mutual}
Zhenting Qi, Mingyuan Ma, Jiahang Xu, Li~Lyna Zhang, Fan Yang, and Mao Yang.
\newblock Mutual reasoning makes smaller llms stronger problem-solvers.
\newblock \emph{arXiv preprint arXiv:2408.06195}, 2024.

\bibitem[Guan et~al.(2025)Guan, Zhang, Liu, Shang, Sun, Zhu, Yang, and Yang]{guan2025rstar}
Xinyu Guan, Li~Lyna Zhang, Yifei Liu, Ning Shang, Youran Sun, Yi~Zhu, Fan Yang, and Mao Yang.
\newblock rstar-math: Small llms can master math reasoning with self-evolved deep thinking.
\newblock \emph{arXiv preprint arXiv:2501.04519}, 2025.

\bibitem[Lightman et~al.(2023)Lightman, Kosaraju, Burda, Edwards, Baker, Lee, Leike, Schulman, Sutskever, and Cobbe]{lightman2023let}
Hunter Lightman, Vineet Kosaraju, Yuri Burda, Harrison Edwards, Bowen Baker, Teddy Lee, Jan Leike, John Schulman, Ilya Sutskever, and Karl Cobbe.
\newblock Let's verify step by step.
\newblock In \emph{The Twelfth International Conference on Learning Representations}, 2023.

\bibitem[He et~al.(2025)He, Liang, Xu, Liu, Chen, Wang, Song, Yu, Liang, Wang, et~al.]{he2025deepmath}
Zhiwei He, Tian Liang, Jiahao Xu, Qiuzhi Liu, Xingyu Chen, Yue Wang, Linfeng Song, Dian Yu, Zhenwen Liang, Wenxuan Wang, et~al.
\newblock Deepmath-103k: A large-scale, challenging, decontaminated, and verifiable mathematical dataset for advancing reasoning.
\newblock \emph{arXiv preprint arXiv:2504.11456}, 2025.

\bibitem[Muennighoff et~al.(2025)Muennighoff, Yang, Shi, Li, Fei-Fei, Hajishirzi, Zettlemoyer, Liang, Cand{\`e}s, and Hashimoto]{muennighoff2025s1}
Niklas Muennighoff, Zitong Yang, Weijia Shi, Xiang~Lisa Li, Li~Fei-Fei, Hannaneh Hajishirzi, Luke Zettlemoyer, Percy Liang, Emmanuel Cand{\`e}s, and Tatsunori Hashimoto.
\newblock s1: Simple test-time scaling.
\newblock \emph{arXiv preprint arXiv:2501.19393}, 2025.

\bibitem[Xia et~al.(2025)Xia, Li, Leong, Wang, and Li]{xia2025tokenskip}
Heming Xia, Yongqi Li, Chak~Tou Leong, Wenjie Wang, and Wenjie Li.
\newblock Tokenskip: Controllable chain-of-thought compression in llms.
\newblock \emph{arXiv preprint arXiv:2502.12067}, 2025.

\bibitem[Shen et~al.(2025)Shen, Huang, Zhao, Liu, Zheng, and Zhu]{shen2025long}
Si~Shen, Fei Huang, Zhixiao Zhao, Chang Liu, Tiansheng Zheng, and Danhao Zhu.
\newblock Long is more important than difficult for training reasoning models.
\newblock \emph{arXiv preprint arXiv:2503.18069}, 2025.

\end{thebibliography}
